\documentclass[aps,pra,notitlepage,twocolumn,superscriptaddress,nobalancelastpage]{revtex4-1}

\usepackage[T1]{fontenc}
\usepackage{amsmath,amssymb,amsfonts,multirow,xcolor,graphicx,float,siunitx}
\usepackage[colorlinks,breaklinks]{hyperref}
\usepackage[caption=false]{subfig}
\definecolor{darkred}{rgb}{0.5,0,0}
\definecolor{darkgreen}{rgb}{0,0.5,0}
\definecolor{darkblue}{rgb}{0,0,0.5}
\hypersetup{linkcolor=darkred,citecolor=darkgreen,urlcolor=darkblue}

\def\ket#1{|#1\rangle}

\def\ketbra#1{|#1\rangle\langle#1|}
\def\bra#1{\langle#1|}
\def\mub#1#2{\ket{\varphi_{#2}^{#1}}}
\def\pmub#1#2{\ketbra{\varphi_{#2}^{#1}}}
\renewcommand{\leq}{\leqslant}
\renewcommand{\geq}{\geqslant}

\newtheorem{definition}{Definition}
\DeclareMathOperator{\Tr}{Tr}

\begin{document}

\title{Genuine high-dimensional quantum steering}
\author{S{\'e}bastien Designolle}
\affiliation{Department of Applied Physics, University of Geneva, 1211 Geneva, Switzerland}
\author{Vatshal Srivastav}
\affiliation{Institute of Photonics and Quantum Sciences, Heriot-Watt University, Edinburgh EH14 4AS, United Kingdom}
\author{Roope Uola}
\affiliation{Department of Applied Physics, University of Geneva, 1211 Geneva, Switzerland}
\author{\\Natalia Herrera Valencia}
\affiliation{Institute of Photonics and Quantum Sciences, Heriot-Watt University, Edinburgh EH14 4AS, United Kingdom}
\author{Will McCutcheon}
\affiliation{Institute of Photonics and Quantum Sciences, Heriot-Watt University, Edinburgh EH14 4AS, United Kingdom}
\author{Mehul Malik}
\affiliation{Institute of Photonics and Quantum Sciences, Heriot-Watt University, Edinburgh EH14 4AS, United Kingdom}
\author{Nicolas Brunner}
\affiliation{Department of Applied Physics, University of Geneva, 1211 Geneva, Switzerland}
\date{\today}

\begin{abstract}
  High-dimensional quantum entanglement can give rise to stronger forms of nonlocal correlations compared to qubit systems, offering significant advantages for quantum information processing.
  Certifying these stronger correlations, however, remains an important challenge, in particular in an experimental setting.
  Here we theoretically formalise and experimentally demonstrate a notion of genuine high-dimensional quantum steering.
  We show that high-dimensional entanglement, as quantified by the Schmidt number, can lead to a stronger form of steering, provably impossible to obtain via entanglement in lower dimensions.
  Exploiting the connection between steering and incompatibility of quantum measurements, we derive simple two-setting steering inequalities, the violation of which guarantees the presence of genuine high-dimensional steering, and hence certifies a lower bound on the Schmidt number in a one-sided device-independent setting.
  We report the experimental violation of these inequalities using macro-pixel photon-pair entanglement certifying genuine high-dimensional steering.
  In particular, using an entangled state in dimension $d=31$, our data certifies a minimum Schmidt number of $n=15$.
\end{abstract}

\maketitle

\textit{Introduction.---}
The possibility of having entanglement between quantum systems with a large number of degrees of freedom opens interesting perspectives in quantum information science \cite{FVMH19}.
In particular, high-dimensional quantum systems can lead to stronger forms of correlations \cite{VWZ02,TAZG04}, featuring increased resilience to noise and losses~\cite{VPB10,EBB+19,ZTV+21}.
This makes them a promising alternative to qubits for applications in quantum technology, in particular for quantum communications~\cite{CBKG02,SS10,MDG+13,MMO+15,ILC+17,CDBO19}.
Experimentally, impressive progress has been achieved in recent years towards the generation and manipulation of high-dimensional entanglement~\cite{DLB+11,MEH+16,KRR+17,WPD+18,VGM+20}.
A key problem is then to certify and characterise this entanglement.
This is challenging not only due to the large number of parameters in the Hilbert space, but also because experimentally available data is typically limited.
Nevertheless, significant progress has been reported in scenarios assuming fully characterised measurement devices~\cite{MGT+17,TDC+17,BVK+18,DMP+18,GHL+18,STF+19,VSP+20}.

It turns out that quantum theory allows one to certify high-dimensional entanglement, as quantified by the notion of Schmidt number ~\cite{TH00,SBL01} (see below), based only on the nonlocal correlations it produces, hence relaxing the requirement of a perfectly calibrated or trusted measurement device.
That is, given some observed data, one can in principle certify the presence of high-dimensional entanglement (i.e., infer a lower bound on the Schmidt number) without making any assumptions about the workings of the measurement devices used.
Beyond their fundamental interest, such black-box tests are also relevant for device-independent quantum information processing~\cite{HP13,SC18,GCH+19}.
Previous works have discussed these questions for Bell nonlocality mostly on the theoretical level~\cite{BPA+08,WYBS17}, with proof-of-principle experiments certifying entangled states of Schmidt number $n=3$ ~\cite{WPD+18} and $n=4$~\cite{CBRS16}.
The experimental certification of higher-dimensional entanglement via nonlocality is extremely demanding technologically, requiring very high state fidelities and offering extremely low tolerance to noise.

In this work, we address these questions from the point of view of quantum steering, a form of quantum correlations intermediate between entanglement and Bell nonlocality~\cite{WJD07,QVC+15}.
Quantum steering relaxes the strict technological requirements of Bell nonlocality by assuming an uncharacterised or untrusted measurement device only on one side.
However, steering tests developed so far can only witness the presence of entanglement (i.e., certify a Schmidt number $n>1$) but do not characterize the entanglement dimensionality~\cite{CS16b,UCNG20,RDB+09}.
Here we develop a notion of genuine high-dimensional quantum steering.
This leads to effective methods for certifying a minimal entanglement dimensionality, i.e., a lower bound on the Schmidt number $n$, in a one-sided device-independent setting.We demonstrate this experimentally with photon pairs entangled in their discretised transverse position-momentum and certify Schmidt numbers up to $n=15$.

\begin{figure*}[ht!]
  \centering
  \subfloat[\label{sfig:steering}]{\includegraphics[width=0.4\textwidth]{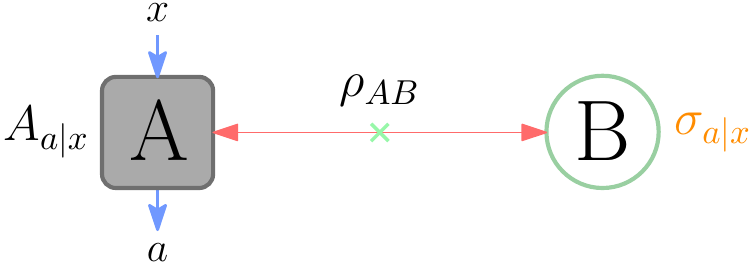}}
  \hspace{0.1\textwidth}
  \subfloat[\label{sfig:dolls}]{\includegraphics[width=0.45\textwidth]{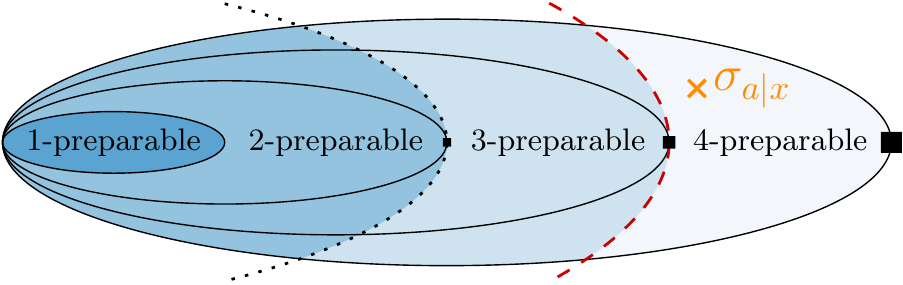}}
  \caption{
    {\bf High-dimensional quantum steering.}
    (a) Alice and Bob share an entangled state $\rho_{AB}$.
    By performing local measurements $A_{a|x}$, Alice remotely steers Bob's subsystem, described by the assemblage $\sigma_{a|x}$.
    (b) As the entanglement dimensionality (the Schmidt number $n$) of the state $\rho_{AB}$ increases, stronger correlations can be created.
    More precisely, by performing well-chosen measurements, Alice can generate for Bob an assemblage $\sigma_{a|x}$ that can provably not have been obtained via any lower-dimensional entangled state.
    To prove this, we define the notion of $n$-preparable assemblages, i.e., that can be produced via $\rho_{AB}$ with Schmidt number $n$.
    This leads to a hierarchy of sets, shown here for $n\leq4$.
    First, one-preparable assemblages (with $\mathrm{SR}(\sigma_{a|x})=0$ thus $\delta(\sigma_{a|x})=1$) feature no quantum steering.
    Next, the two- and three-preparable sets contain assemblages achievable with entangled states of Schmidt number $n=2$ and $n=3$, respectively.
    Beyond this, there exist assemblages that are not three-preparable, hence featuring genuine four-dimensional steering, as witnessed by violation of a steering inequality (red dashed line corresponding to $\delta(\sigma_{a|x})>3$).
    This guarantees the presence of an entangled state of Schmidt number $n=4$ in a one-sided device-independent setting.
  }
  \label{fig:hds}
\end{figure*}

Consider a scenario featuring two distant parties, Alice and Bob, sharing a bipartite quantum state $\rho_{AB}$.
In the task of steering (see Fig.~\ref{sfig:steering}), Alice performs several possible quantum measurements on her subsystem, thus remotely ``steering'' the state of Bob's subsystem to
\begin{equation}\label{eqn:assemblage}
  \sigma_{a|x} = \Tr_A\big[ (A_{a|x} \otimes \openone_B) \rho_{AB}\big],
\end{equation}
where $x$ denotes Alice's choice of measurement and $a$ its outcome.
Alice's measurements are represented by a set of positive operators $A_{a|x}$ satisfying ${\sum_a A_{a|x}= \openone_A}$ for all $x$.
The collection $\{\sigma_{a|x} \}_{a,x}$ of the possible (unnormalised) steered states is termed an assemblage, referred to as $\sigma_{a|x}$ in the following.
When this assemblage can be produced without the use of entanglement, i.e., via a so-called local hidden state (LHS) model \cite{WJD07}, the assemblage is called unsteerable.
If this is not possible, then the assemblage demonstrates steering.
This effect has been investigated experimentally, mostly with qubit entanglement~\cite{SJWP10,SGA+10,WRS+12,BES+12}.

Here we are specifically interested in the situation where Alice and Bob share a quantum state featuring high-dimensional entanglement.
Consider for instance a $d\times d$ maximally entangled state $\ket{\phi_d} = \sum_{j=0}^{d-1} \ket{j,j} / \sqrt{d}$.
By using any set of incompatible measurements, Alice can generate on Bob's side an assemblage featuring steering~\cite{QVB14,UMG14}.
Moreover, for large dimensions, the robustness to noise and losses of these assemblages is known to increase~\cite{SC15,MRY+15,ZWLZ18,DSFB19}.
This suggests that high-dimensional entanglement can in fact lead to assemblages featuring a stronger form of quantum correlations.
In particular, by using well-chosen measurements, Alice may generate an assemblage for Bob that could not have been created using lower-dimensional entanglement.
Below, we formalise this intuition and define the notion of genuine high-dimensional steering.

Specifically, we characterise the entanglement dimensionality through the concept of Schmidt number~\cite{TH00,SBL01}.
The Schmidt number of a state $\rho_{AB}$ is the minimum $n$ such that there exists a decomposition ${\rho_{AB}=\sum_j p_j \ket{\psi_j}\bra{\psi_j}}$
where all $\ket{\psi_j}$ are pure entangled states of Schmidt rank at most $n$.
For pure states, the Schmidt number is simply equal to the Schmidt rank.
This motivates us to define the notion of $n$-preparable assemblages.

\begin{definition}
  An assemblage $\sigma_{a|x}$ acting on $\mathbb{C}^d$ is $n$-preparable, with $1\leq n\leq d$, when it can be decomposed as $\sigma_{a|x} = \Tr_A\big[ (A_{a|x} \otimes \openone_B) \rho_{AB}\big]$ where $\rho_{AB}\in \mathcal{D}(\mathcal{H}^A \otimes \mathcal{H}^B)$ has Schmidt number $n$ and $d$ is the dimension of the Hilbert space $ \mathcal{H}^B\equiv \mathbb{C}^d$.
\end{definition}

In other words, an $n$-preparable assemblage $\sigma_{a|x}$ can be prepared via suitable operations on an entangled state of Schmidt number $n$.
However, one could also prepare this assemblage via operations on a state with a larger Schmidt number.
This implies that any $n$-preparable assemblage is also straightforwardly $(n+1)$-preparable, which leads to a nested structure of assemblages, as shown in Fig.~\ref{sfig:dolls}.
For $n=1$ we recover the usual definition of steering: any one-preparable assemblage can be reproduced via a LHS model \cite{WJD07} or equivalently via a separable state (i.e., with Schmidt number $n=1$) of arbitrary dimension on Alice's side~\cite{KSC+15,MGH+16}.
On the other hand, for $n=d$ we obtain the full set of quantum assemblages on $\mathbb{C}^d$, as any decomposition of a density matrix can be remotely generated via shared entanglement and well-chosen local measurements~\cite{Gis84,HJW93}.

\begin{figure*}[ht!]
  \centering
  \subfloat[\label{sfig:setup}]{\includegraphics[width=8cm]{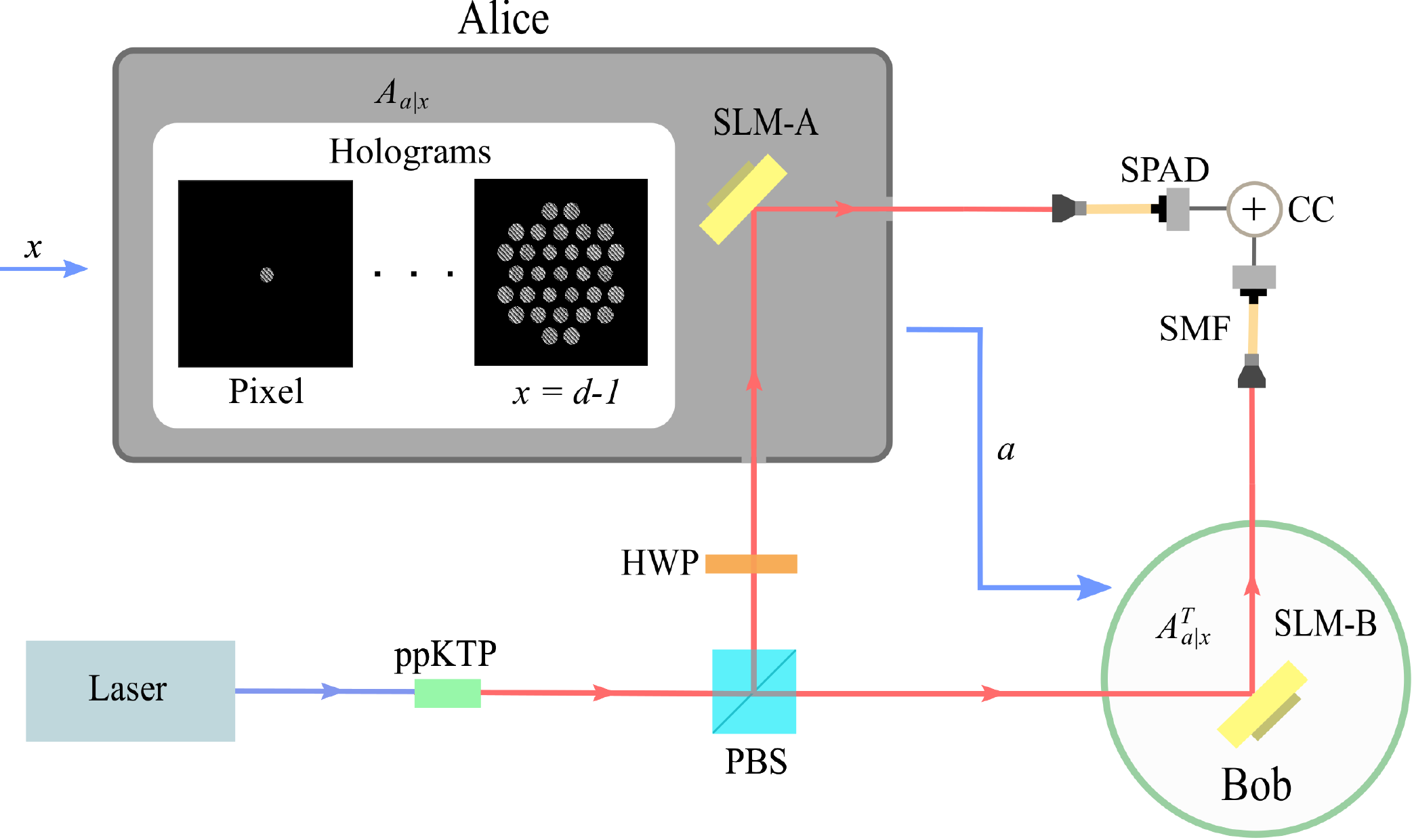}}
  \hspace{1cm}
  \raisebox{0.7cm}{\subfloat[\label{sfig:coincidences}]{\includegraphics[width=7cm]{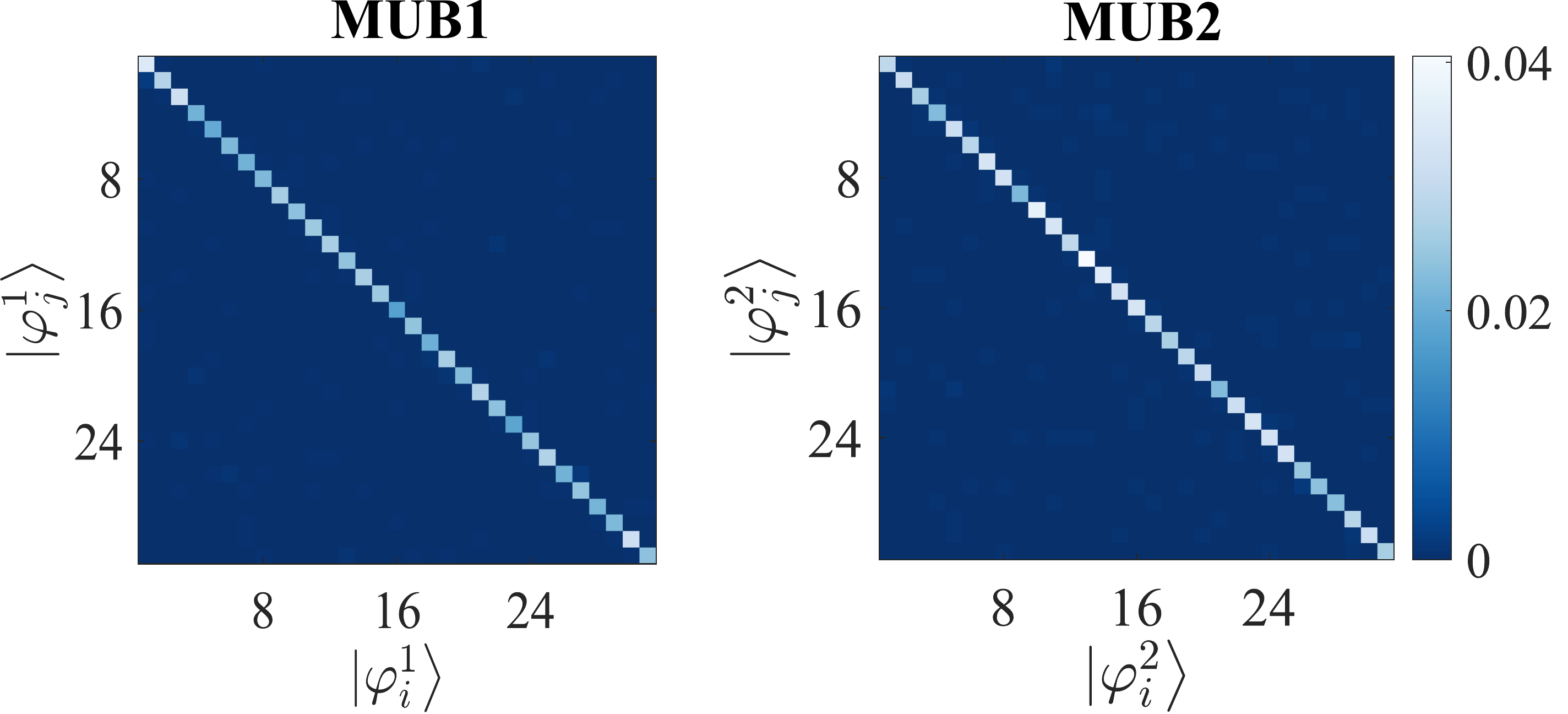}}}
  \caption{
    \textbf{Experimental realisation.}
    (a) One photon from an entangled photon pair and a classical bit $x$ are distributed to the untrusted party, Alice.
    She generates holograms on a spatial light modulator (SLM-A) to perform the projection $A_{a|x}$ onto outcome $a$ for the given basis $x$ and passes $a$ to the trusted party, Bob.
    Bob receives the other photon along with this classical information, forming the conditional state $\sigma_{a|x}$, and he performs the projection $A^T_{a|x}$ according to his chosen steering inequality.
    Coincident photon detection events are then used to evaluate the steering inequality, under the fair sampling assumption, allowing us to certify genuine high-dimensional steering.
    (b) Normalised two-photon coincidence counts in a pair of $d=31$ dimensional mutually unbiased pixel bases ($x=1$ and $2$).
    Using these measurements we obtain the maximum value of $\delta(\sigma_{a|x}) \geq 14.1 \pm 0.6$ that demonstrates genuine 15-dimensional steering (i.e., Schmidt number $n=15$).
  }
  \label{fig:setup}
\end{figure*}

Interestingly, there exist assemblages that are $n$-preparable without being {$(n-1)$}-preparable, thus featuring genuine $n$-dimensional steering.
For instance, in the case $n=4$, there are assemblages that cannot be created by only using entangled states with Schmidt number 3.
Such assemblages thus feature genuine four-dimensional steering and guarantee that the underlying states have Schmidt number $n=4$ (see Fig.~\ref{sfig:dolls}).
It turns out, however, that the general characterisation of the set of $n$-preparable assemblages is challenging.
Notably, standard methods that allow for a full characterisation of unsteerable assemblages do not work here: determining whether an assemblage is one-preparable can be cast as a semidefinite programming~\cite{CS16b}, which does not appear to be the case for $n$-preparability when $n>1$.

Nevertheless, we can derive a necessary criterion for $n$-preparability in the case where Alice has two possible measurements.
We use the notion of steering robustness ($\mathrm{SR}$), a convex quantifier of quantum steering~\cite{PW15,CS16b}:
\begin{equation}
  \label{eqn:SRmain}
  \!\!\!\mathrm{SR}(\sigma_{a|x}) = \!\min_{t,\tau_{a|x}}\Bigg\{t\geq 0\,\bigg\vert\, \frac{\sigma_{a|x} + t{\tau_{a|x}}}{1+t}\text{ unsteerable}\Bigg\},
\end{equation}
where the minimisation is over all assemblages $\tau_{a|x}$ with the same dimension and numbers of inputs and outputs as $\sigma_{a|x}$.
The SR quantifies the robustness of $\sigma_{a|x}$ to an arbitrary noise $\tau_{a|x}$ before becoming unsteerable.
Specifically, since any assemblage $\sigma_{a|x}$ that is $n$-preparable can by definition be written as in Eq.~\eqref{eqn:assemblage} where $\rho_{AB}$ has Schmidt number $n$, we can use the convexity of the steering robustness to upper bound $\mathrm{SR}(\sigma_{a|x})$ by
\begin{align}
  &\sum_j p_j\,\mathrm{SR}\Big\{\Tr_A\big[ (A_{a|x} \otimes \openone_B) \ket{\psi_j}\bra{\psi_j}\big]\Big\}\\
  \leq&\max_{\ket{\psi}\in\mathbb{C}^n\otimes\mathbb{C}^n}\mathrm{SR}\Big\{\Tr_A\big[ (A_{a|x} \otimes \openone_B) \ket{\psi}\bra{\psi}\big]\Big\}.
\end{align}
Importantly, the assemblages relating to this last upper bound act on $\mathbb C^n$.
Using the connection between quantum steering and measurement incompatibility~\cite{QVB14,UMG14,UBGP15,KBUP17} and a recent result identifying the most incompatible pairs of quantum measurements in a given dimension~\cite{DFK19}, we find that the steering robustness of any $n$-preparable assemblage is upper bounded by $(\sqrt{n}-1)/(\sqrt{n}+1)$; all details are in Appendix~\ref{app:hd}.
Hence, any $n$-preparable assemblage $\sigma_{a|x}$ (with two inputs for Alice) satisfies
\begin{equation}\label{eqn:ceiling}
  n\geq\left(\frac{1+\mathrm{SR}(\sigma_{a|x})}{1-\mathrm{SR}(\sigma_{a|x})}\right)^2\equiv\delta(\sigma_{a|x}),
\end{equation}
so that violating this inequality amounts to certifying that the assemblage is not $n$-preparable, i.e., the shared state has at least a Schmidt number of $n+1$.

The inequality~\eqref{eqn:ceiling} turns out to be tight for all $n\leq d$: if $\rho_{AB}$ is a maximally entangled state of dimension $n \times n$, and Alice performs projective measurements onto two mutually unbiased bases (MUBs), then the resulting assemblage (embedded in $\mathbb{C}^d$) saturates the bound.
Recall that two orthonormal bases are called mutually unbiased if the scalar product of any vector from the first basis with any vector from the second basis is equal to $1/\sqrt{n}$.
Below, we use a standard construction of sets of MUBs when $d$ is prime (see~Appendix~\ref{app:mub}).

From the above, in particular inequality~\eqref{eqn:ceiling}, we see that given the steering robustness of an assemblage, we obtain a lower bound on the dimension $n$ such that $\sigma_{a|x}$ is $n$-preparable.
Full tomography on Bob's side would make the exact computation of the steering robustness possible via semidefinite programming~\cite{CS16b}.
A more effective (and experimentally friendly) method consists in using the results of Refs~\cite{UKS+19,ULMH16} showing that the steering robustness can be lower bounded via a steering functional involving only a pair of MUBs measurements (denoted $M_{a|x}$) for both Alice and Bob, namely
\begin{equation}\label{eqn:steering-ineq}
  \frac{1}{\lambda}\sum_{a,x}\Tr\big[(M_{a|x}\otimes M_{a|x}^T)\rho_{AB}\big]-1\leq\mathrm{SR}(\sigma_{a|x})\leq\frac{\sqrt{n}-1}{\sqrt{n}+1},
\end{equation}
where $\lambda=1+1/\sqrt{d}$  (see~Appendix~\ref{app:ineq}).
The second inequality in Eq.~\eqref{eqn:steering-ineq} corresponds to the result of Eq.~\eqref{eqn:ceiling} and gives a steering inequality valid for all $n$-preparable assemblages.
For $n = 1$, we recover the inequality ($\sum_{a,x}\Tr\big[(M_{a|x}\otimes M_{a|x}^T)\rho_{AB}\big] \leq 1+1/\sqrt{d}$), the violation of which certifies the presence of entanglement (i.e., Schmidt number $n>1$)~\cite{WPD+18,CS16b,UCNG20,RDB+09}.
Clearly, the inequality \eqref{eqn:steering-ineq} is more general, and provides a lower bound on the Schmidt number $n$ depending on the amount of violation.

Note that this steering inequality as well as the relation \eqref{eqn:ceiling} apply to general assemblages (involving, e.g., mixed states).
Note also that the inequality can be saturated by using a $d \times d$ pure maximally entangled state and projective measurements onto MUBs.
In~Appendix~\ref{app:ineq} we derive the critical noise threshold for violating the inequality for isotropic states (mixture of a maximally entangled state with white noise).
The above method is well-adapted to experiments and allows us to certify genuine high-dimensional steering in practice.
We use photon pairs entangled in their discrete transverse position-momentum, also known as ``pixel'' entanglement~\cite{VSP+20}.
This platform allows us to access generalised $d$-dimensional measurements with a very high quality in dimensions up to $d=31$.
As shown in Fig.~\ref{sfig:setup}, a nonlinear ppKTP crystal is pumped with a continuous-wave ultraviolet laser (\SI{405}{\nano\meter}) to produce a pair of pixel-entangled infrared photons (\SI{810}{\nano\meter}) via type-II spontaneous parametric down-conversion (SPDC).
The photon pairs are separated by a polarising beam splitter (PBS) and directed to Alice and Bob, who each have access to a holographic spatial light modulator (SLM) for performing generalised projective measurements in the pixel basis or any of its MUBs.
The holograms used for performing projective measurements are optimised by tailoring the size and spacing of the pixels based on the knowledge of the joint-transverse-momentum-amplitude (JTMA) of the generated biphoton state~\cite{SH16}.
This choice of basis warrants that the state well approximates a maximally entangled state and that, in addition to the strong correlations in the pixel basis, pixel-MUB are also strongly correlated owing to momentum conservation of the narrow-band, weakly focused pump.
The SLM holograms ensure that only photons carrying pixel/pixel-MUB modes of interest couple efficiently to single-mode fibres (SMF) and are subsequently detected by single-photon avalanche detectors (SPADs).
This allows us to reconstruct the results of the measurement operators $M_{a|x}$ and $M_{a|x}^T$, and implement them in the steering inequality~\eqref{eqn:steering-ineq}.

\begin{figure}[hb!]
  \centering
  \includegraphics[width=\columnwidth]{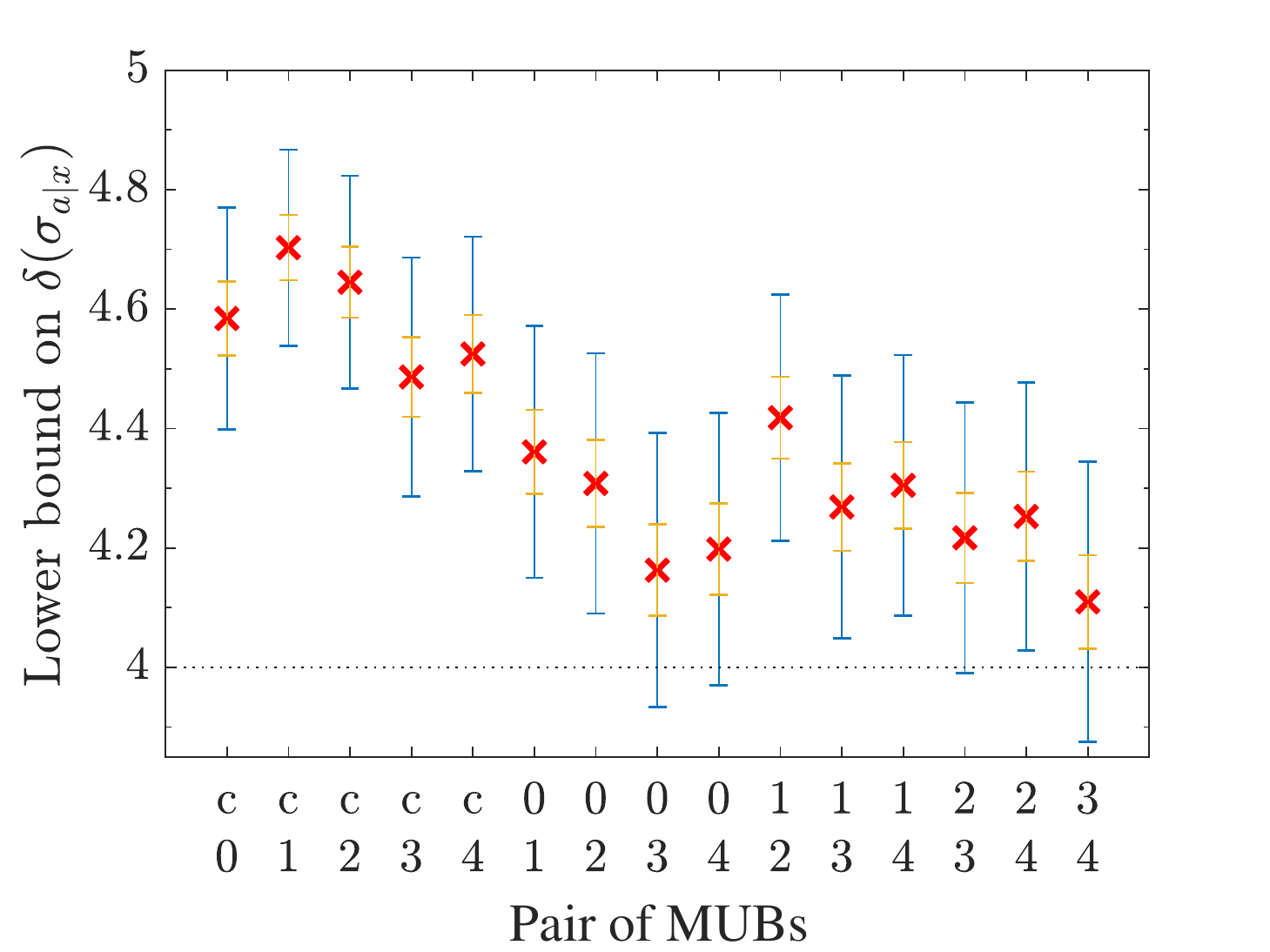}
  \caption{
    {\bf Experimental certification of genuine five-dimensional steering.}
    A photon pair with entanglement in dimension $d=5$ is generated.
    The complete set of MUBs features $d+1=6$ bases, labelled by `c' for computational and $0\ldots4$, which leads to 15 possible pairs of MUBs to be measured by both Alice and Bob.
    For each pair, the certified dimension is given by the ceiling of the quantity $\delta(\sigma_{a|x})$; the steering robustness being estimated via the steering inequality~\eqref{eqn:steering-ineq}.
    Here all pairs of MUBs certify the presence of genuine five-dimensional steering, and hence maximal Schmidt number $n=d=5$, within one standard deviation.
    The second error bar represents three standard deviations.
  }
  \label{fig:res5}
\end{figure}

It is important to note that the measurements actually performed in the experiment only have two outcomes, depending on whether the photon detector clicks or not.
While it is common practice to reconstruct full projective measurements out of these dichotomic ones~\cite{DLB+11,CBRS16,WPD+18,ZWLZ18}, the underlying assumption is strong since the corresponding steering scenarios are inherently different, having a different number of inputs and outputs.
Note also that due to detector and system inefficiencies (see~Appendix~\ref{app:exp}) we are working under the fair-sampling hypothesis; however, no subtraction of background or accidental counts is performed.
\begin{table}[!b]
  \centering
  \begin{tabular}{|c||c|c|c|}
    \hline
    Dimension & \multicolumn{2}{c|}{Lower bound on $\delta(\sigma_{a|x})$} & Certified Schmidt \\
    $d$       &         ~~~Minimum~~~        &        ~~~Maximum~~~        &  number $n$       \\ \hline \hline
    5         &         $ 4.1\pm0.1$         &        $ 4.7\pm0.1$         &        5          \\
    7         &         $ 5.1\pm0.2$         &        $ 6.4\pm0.1$         &        7          \\
    11        &         $ 6.3\pm0.3$         &        $ 9.1\pm0.2$         &        10         \\
    13        &         $ 7.0\pm0.3$         &        $10.1\pm0.3$         &        11         \\
    17        &         $ 9.3\pm0.3$         &        $12.4\pm0.3$         &        13         \\
    19        &         $10.1\pm0.5$         &        $13.6\pm0.5$         &        14         \\ \hline
    23        &              \multicolumn{2}{c|}{$11.4\pm0.5$}             &        12         \\
    29        &              \multicolumn{2}{c|}{$12.1\pm0.6$}             &        13         \\
    31        &              \multicolumn{2}{c|}{$14.1\pm0.6$}             &        15         \\
    \hline
  \end{tabular}
  \vspace{5pt}
  \caption{
    {\bf Experimental results for higher dimensions.}
    Entanglement is prepared in prime dimensions $d$ from 5 to 31.
    For each $d$, we provide the minimum and maximum values of the quantity $\delta(\sigma_{a|x})$, the ceiling of which gives a lower bound on the certified Schmidt number $n$.
    For $d=5$, these values correspond to those of Fig.~\ref{fig:res5}.
    For $d=19$, a Schmidt number of $n=14$ can be certified (for the best pair of MUBs), while all 190 possible pairs certify (at least) $n=11$.
    Moreover, for $d=31$, the data certifies a Schmidt number $n=15$, i.e., genuine 15-dimensional quantum steering.
    Note that for $d \geq 23$, the time required for measuring all $d+1$ MUBs scales unfavourably (in particular for the computational basis), thus only one pair of MUBs was measured.
    In higher dimensions, the errors are larger due to higher count rates.
  }
  \label{tab:reshigh}
\end{table}

The results are given in Fig.~\ref{fig:res5} and Table~\ref{tab:reshigh}.
Note that since there are $d+1$ MUBs in (prime) dimensions $d$, there are $d(d+1)/2 $ possible pairs of them, giving rise to potentially different certified dimensions.
For $d=5$, we consider all 15 possible pairs of MUBs, for all of which we find $\delta(\sigma_{a|x})>4$ (see Fig.~\ref{fig:res5}), thus certifying genuine five-dimensional steering (i.e., Schmidt number $n=5$).
That is, none of this data could be reproduced with entangled states of Schmidt number $n\leq 4$.
Of all possible pairs, those utilising the pixel basis (also referred to as computational or simply~`c') exhibit slightly better bounds owing to the higher visibility in this basis, since it is the natural Schmidt basis resulting from momentum conservation.

Next we investigate higher dimensions, up to $d=31$.
Note that for $d\geq 23$, we measured only one pair of MUBs to optimise the total data acquisition time, as the number of single-outcome measurements required increases with $O(d^2)$.
In Table~\ref{tab:reshigh} we only show, for simplicity, the minimum and maximum values obtained for the parameter $\delta(\sigma_{a|x})$; a Schmidt number of $n=15$ can be certified when using an entangled state in dimension $d=31$.
Moreover, for $d=19$, all 190 possible pairs of MUBs certify (at least) Schmidt number $n=11$ and up to $n=14$.
The total measurement time for measuring two MUBs (excluding the computational basis) was 40 seconds for $d=5$ and 16 minutes for $d=31 $.

We have developed the concept of genuine high-dimensional steering, leading to effective methods for certifying a lower bound on the entanglement dimensionality (the Schmidt number) in a one-sided device-independent setting, as demonstrated in a photonic experiment.
Moreover, our approach can be readily applied to other quantum platforms using different degrees of freedom (see~Appendix~\ref{app:platforms}).
Our work could be of significant interest for information-theoretic tasks such as randomness generation and cryptography.
More generally, this represents an important step towards the realisation of noise-robust, high-capacity quantum networks in the near future.

\section*{Acknowledgements}

S.D., V.S., and R.U.~contributed equally to this work.
The authors thank Marcus Huber and Paul Skrzypczyk for discussions.
N.B., S.D., and R.U.~acknowledge financial support from the Swiss National Science Foundation (Starting grant DIAQ and NCCR QSIT).
M.M., N.H.V., W.M., and V.S.~acknowledge financial support from the QuantERA ERA-NET Co-fund (FWF Project I3773-N36) and the UK Engineering and Physical Sciences Research Council (EPSRC) (EP/P024114/1).

\bibliography{DSU+21}

\bibliographystyle{sd2}

\appendix

\onecolumngrid
\section*{Appendix}
\twocolumngrid

\subsection{Dimension certificate}
\label{app:hd}

In a steering scenario (see Fig.~\ref{sfig:steering}), upon measurement of $A_{a|x}$ on Alice's side, the state assemblage created on Bob's side is ${\sigma_{a|x}=\Tr_A[(A_{a|x}\otimes\openone)\rho_{AB}]}$.
The assemblage is said to be unsteerable if there exists a collection of (unnormalised) local states $\{\rho_\mu\}_\mu$ on Bob's system such that
\begin{equation}\label{eqn:lhs}
  \sigma_{a|x}=\sum_\mu D_\mu(a|x)\rho_\mu,
\end{equation}
where $D_\mu(a|x)$ are deterministic post-processings.
When no such local hidden state model exists, the assemblage is called steerable~\cite{CS16b,UCNG20}.

The steering robustness measures how much a state assemblage can tolerate general noise before becoming unsteerable~\cite{PW15}.
Formally,
\begin{equation}\label{eqn:sr}
  \!\!\!\mathrm{SR}(\sigma_{a|x}) = \!\min_{t,\tau_{a|x}}\Bigg\{t\geq 0\,\bigg\vert\, \frac{\sigma_{a|x} + t{\tau_{a|x}}}{1+t}\text{ unsteerable}\Bigg\},
\end{equation}
where the minimisation is over all state assemblages $\tau_{a|x}$ having the same numbers of inputs and outputs as $\sigma_{a|x}$.
For further reference we also define the consistent steering robustness $\mathrm{CSR}(\sigma_{a|x})$ by only allowing mixing with assemblages that have the same total state as $\sigma_{a|x}$, i.e., $\sum_a\sigma_{a|x}=\sum_a\tau_{a|x}$, the value of $x$ being irrelevant due to no-signalling.

Suppose that the state assemblage $\sigma_{a|x}$ admits a decomposition of the form
\begin{equation}\label{eqn:decomp}
  \sigma_{a|x}=\sum_j p_j\tau_{a|x}^{(j)},
\end{equation}
where the state assemblages $\tau_{a|x}^{(j)}$ all have dimension at most $n$.
Such an assemblage would occur when a steering experiment is carried out with a state $\rho_{AB}$ having a Schmidt number at most $n$.

To develop our criterion for high-dimensional steering, we note that assuming that Eq.~(\ref{eqn:decomp}) holds for a given $n$ results in an upper bound on the steering robustness due to its convexity (see, e.g., Ref.~\cite{CS16b})
\begin{equation}
  \mathrm{SR}(\sigma_{a|x})\leq\sum_j p_j\mathrm{SR}\left(\tau_{a|x}^{(j)}\right).
\end{equation}
Next we can use the well-known ordering between the steering robustness and the consistent steering robustness (see, e.g., Ref.~\cite[Eq.~(32)]{CS16a}) in order to get
\begin{equation}
  \mathrm{SR}(\sigma_{a|x})\leq\sum_j p_j\mathrm{CSR}\left(\tau_{a|x}^{(j)}\right).
\end{equation}
At this point we take advantage of the connection between steering and joint measurability (see, e.g., Ref.~\cite{UBGP15}) to obtain
\begin{equation}\label{eqn:csr-ir}
  \!\!\!\mathrm{CSR}\left(\tau_{a|x}^{(j)}\right)=\mathrm{IR}\left(\tau_{B,j}^{-\frac12}\,\tau_{a|x}^{(j)}\,\tau_{B,j}^{-\frac12}\right)\leq\max_{M_{a|x}}\,\mathrm{IR}(M_{a|x}),
\end{equation}
where $\tau_{B,j}=\sum_a\tau_{a|x}^{(j)}$ is independent of $x$ thanks to no-signalling and $\mathrm{IR}(M_{a|x})$ is the incompatibility robustness, which is the equivalent of the steering robustness of Eq.~\eqref{eqn:sr} for incompatibility of quantum measurements~\cite{CS16a}.
Note that the initial assumption on the dimension of the assemblage $\tau_{a|x}^{(j)}$ translates into the constraint that the measurements $M_{a|x}$ over which the maximisation is performed in Eq.~\eqref{eqn:csr-ir} should have dimension at most $n$.

Here come into the play recent results on the most incompatible pairs of measurements~\cite{DFK19}.
Therein, when $x$ can only take two values, the maximum of Eq.~\eqref{eqn:csr-ir} is computed, so that we eventually get
\begin{equation}\label{eqn:most-incompatible}
  \mathrm{SR}(\sigma_{a|x})\leq\frac{\sqrt{n}-1}{\sqrt{n}+1}.
\end{equation}
Importantly, pairs of MUBs saturate the bound, which makes it particularly powerful with them.
Note that Ref.~\cite{DFK19} only provides such a tight bound in the case of pairs of measurements.
For high dimensions, nothing similar exists in the literature for more measurements, up to our knowledge.

All in all, we have proven that if the inequality \eqref{eqn:most-incompatible} is violated, then no decomposition of the form of Eq.~\eqref{eqn:decomp} can exist for a given $n$.
This then means that no lower-dimensional model of steering is responsible for the one observed, or, put differently, that the observed steering phenomenon is genuinely high-dimensional.
Note also that this procedure allows for a one-sided device-independent certification of the Schmidt number of the underlying state.

\subsection{MUBs in prime dimensions}
\label{app:mub}

In prime dimension $d$, the construction of a complete set of $d+1$ MUBs is quite elementary and can be traced back to Ref.~\cite{Iva81}.
The first basis is the computational one, denoted $\{\ket{l}\}_{l=0}^{d-1}$, and the other $d$ bases are $\{\mub{x}{a}\}_{a=0}^{d-1}$, labelled by $x=0\ldots d-1$, where
\begin{align}\label{eqn:mub}
  \mub{x}{a} = \frac{1}{\sqrt{d}}\sum_{l=0}^{d-1}\omega^{al+xl^2}\ket{l},
\end{align}
with $\omega=\exp(2\mathrm{i}\pi/d)$ is a $d$-th root of the unity.

\subsection{Steering inequality}
\label{app:ineq}

A steering inequality enables to certify the steerability of a given assemblage.
Even more, from the amount of violation we can estimate the steering robustness: from the optimisation problem~\eqref{eqn:sr} defining the steering robustness (which is a minimisation), standard methods give a dual problem (which is a maximisation), see, e.g., Ref.~\cite{BV04}.
Then any feasible point of this dual (i.e., a point satisfying the constraints) gives a lower bound on the steering robustness.

This dual is, for instance, stated in Ref.~\cite{UKS+19} and reads
\begin{align}\label{eqn:dual}
  \mathrm{SR}(\sigma_{a|x})=\max_{F_{a|x}}&\quad\sum_{a,x}\Tr(F_{a|x}\sigma_{a|x})-1\\
  \mathrm{s.t.}&\quad\sum_{a,x}\Tr(F_{a|x}\tau_{a|x})\leq 1\quad\forall\,\tau_{a|x}\in \mathrm{US}\nonumber\\
  &\quad F_{a|x}\geq0\quad\forall\,a,x,\nonumber
\end{align}
where $\mathrm{US}$ refers to the set of unsteerable assemblages of the form~\eqref{eqn:lhs}.
Any feasible point $\{F_{a|x}\}$ is a steering inequality, whose amount of violation can be translated into a lower bound on the steering robustness.
It is noteworthy that although we have presented steering on the level of assemblages, requiring tomography on Bob's side to be accessible, steering inequalities given are solely based on correlations between Alice and Bob (see below).

\begin{figure}[hb!]
  \centering
  \includegraphics[width=0.8\columnwidth]{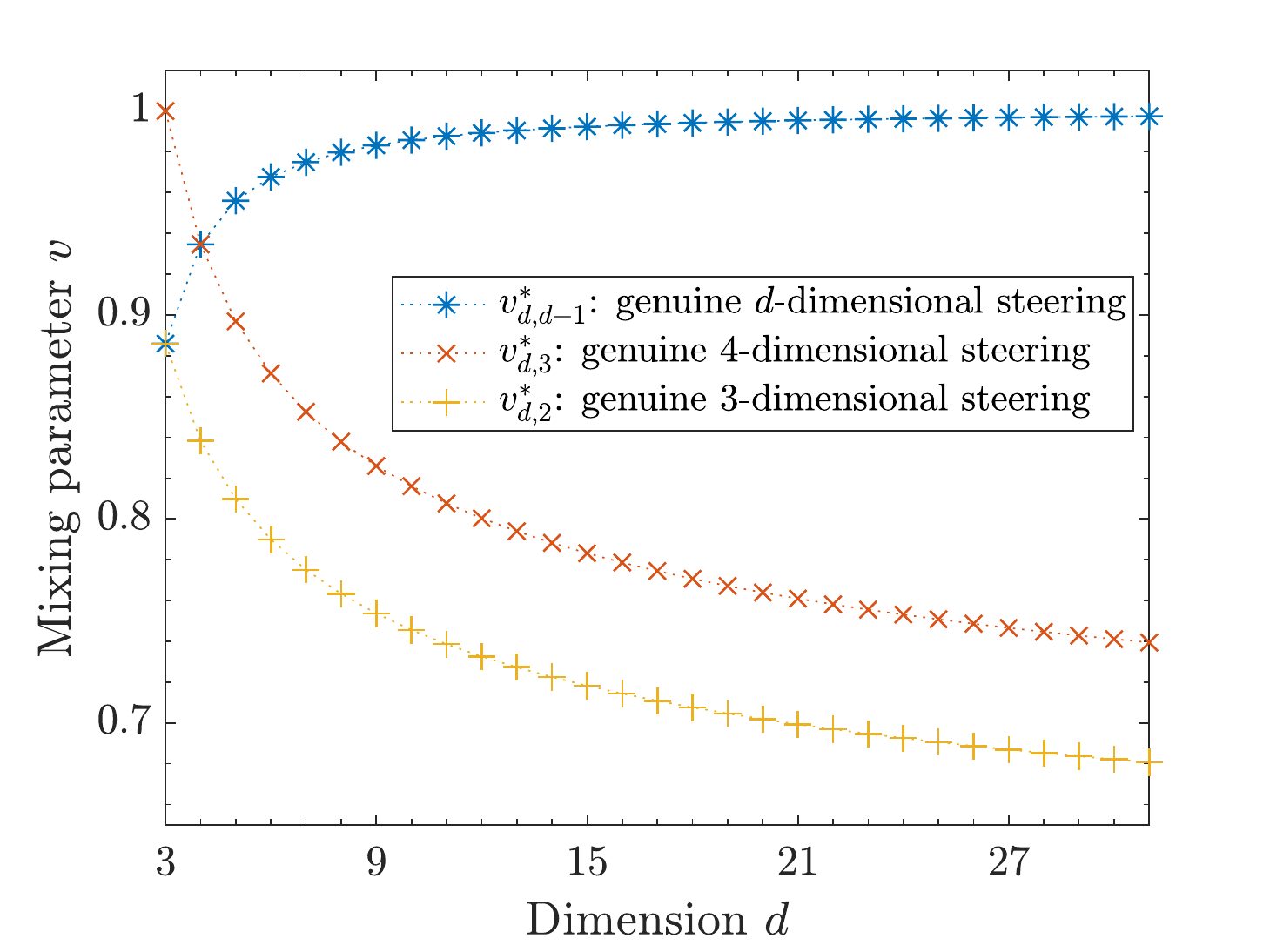}
  \caption{
    Critical mixing parameters to observe genuine high-dimensional steering when performing projective measurements onto two MUBs on a shared isotropic state of local dimension 3 to 31.
    From top to bottom, the three curves correspond to the threshold above which one can certify genuine $d$-dimensional, 4-dimensional, and 3-dimensional steering.
    The corresponding asymptotes are respectively 1, $\sqrt{3}/(1+\sqrt{3})\approx0.6340$, and $\sqrt{2}/(1+\sqrt{2})\approx0.5858$.
    \vspace{-3pt}
  }
  \label{fig:vis}
\end{figure}

In the case of the maximally entangled state, performing the measurements $A_{a|x}$ on Alice's side results in the state assemblage $\sigma_{a|x}=A_{a|x}^T/d$.
Taking inspiration from Ref.~\cite{DSFB19} we propose the following steering inequality:
\begin{equation}\label{eqn:ansatz}
  F_{a|x}=\frac{A_{a|x}^T}{\max\limits_\mu\bigg\|\sum\limits_{a,x}D_\mu(a|x)A_{a|x}\bigg\|_\infty},
\end{equation}
where $\|\cdot\|_\infty$ is the spectral norm and $D_\mu(a|x)$ are deterministic post-processings.
For pairs of MUBs, the denominator is simply ${\lambda=1+1/\sqrt{d}}$~\cite{ULMH16}.
Then the bound~\eqref{eqn:steering-ineq} given in the main text can be recovered
\begin{align}
  \mathrm{SR}(\sigma_{a|x})&\geq\sum_{a,x}\Tr(F_{a|x}\sigma_{a|x})-1\nonumber\\
  &\geq\sum_{a,x}\Tr\left(\frac{A_{a|x}^T}{\lambda}\Tr_A\big[(A_{a|x}\otimes\openone_B)\rho_{AB}\big]\right)-1\nonumber\\
  &\geq\frac{1}{\lambda}\sum_{a,x}\Tr\big[(A_{a|x}\otimes A_{a|x}^T)\rho_{AB}\big]-1.\nonumber
\end{align}
Crucially, this last expression can be evaluated by means of the coincidences measured experimentally when performing $A_{a|x}=\pmub{x}{a}$ on Alice's side and $A_{a|x}^T$ on Bob's side.

As a summary, what can be done experimentally is to compute a lower bound on $\mathrm{SR}(\sigma_{a|x})$ via plugging Eq.~\eqref{eqn:ansatz} in the objective function of Eq.~\eqref{eqn:dual}.
If this upper bound is good enough to violate Eq.~\eqref{eqn:most-incompatible} for some $n\geq2$, then high-dimensional steering can be certified.

As an illustration, when considering the simplest model of a perfect pair of MUB measurements on an isotropic state, that is, a mixture ${v\ketbra{\phi_d}+(1-v)\openone_{d^2}/d^2}$ of the maximally entangled state with white noise, the resulting assemblage will exhibit genuine \mbox{$(n+1)$-dimensional} steering when the mixing parameter satisfies $v>v^*_{d,n}$, where the critical visibility is given by
\begin{equation}
  v^*_{d,n} =\frac{\left(d+\sqrt{d}-1\right)\sqrt{n}-1}{(d-1)\left(\sqrt{n}+1\right)}.
\end{equation}
In Fig.~\ref{fig:vis} we give three typical curves for this critical parameter, making clear that, for fixed $n$, genuine $n$-dimensional steering becomes simpler to demonstrate when $d$ increases.

\subsection{Experimental details}
\label{app:exp}

A continuous-wave grating-stabilised laser at \SI{405}{\nano\meter} (Toptica DL Pro HP) is shaped by a telescope system composed of two lenses with ${f_1 = \SI{250}{\milli\meter}}$ and ${f_2 = \SI{50}{\milli\meter}}$ and is loosely focused onto a periodically poled Potassium Titanyl Phosphate (ppKTP) crystal (${\SI{1}{\milli\meter}\times\SI{2}{\milli\meter}\times\SI{5}{\milli\meter}}$) with a beam waist of approximately \SI{200}{\micro\meter} and a power of \SI{75}{\milli\watt}.
The ppKTP crystal is temperature-tuned with a custom-made resistive oven that keeps it at \SI{30}{\degreeCelsius} to meet the phase-matching conditions for type-II spontaneous parametric down conversion (SPDC) from \SI{405}{\nano\meter} to \SI{810}{\nano\meter}.
This process generates pairs of orthogonally polarised infrared photons entangled in their position-momentum.
The photon pairs are then separated using a polarised beam splitter (PBS) and made incident on two phase-only spatial light modulators (SLMs), which are placed at the Fourier plane of the crystal using a \SI{250}{\milli\meter} lens.

Computer-generated diffractive holograms are displayed on the SLMs (Hamamatsu X10468-02, pixel pitch of \SI{20}{\micro\meter}, resolution of 792$\times$600, diffraction efficiency of 65\% at \SI{810}{\nano\meter}) to perform generalised projective measurements on the incident photons such that only selected modes of the macro-pixel basis or its mutually unbiased bases efficiently couple to single-mode fibres (SMFs).
The SMFs guide the photons selected by the SLMs to single-photon avalanche detectors (Excelitas SPCM-AQRH-14-FC) where they are detected with an efficiency of 60\%.
A coincidence counting logic (UQDevices) records time-coincident events within a window of \SI{0.2}{\nano\second}.
We increase the accuracy of the projective measurements made with the combination of the SLM and the SMF through an intensity-flattening technique~\cite{BVB+18}, where we install carefully designed telescopes (IFTs) on both Alice's and Bob's sides.
The IFT reduces the size of the mode propagating from the SLM to the SMF by a factor of 3.3 in order to increase the collected two-photon modal bandwidth at the expense of a tolerable loss of 5\% on the total efficiency.

The coincidence counts obtained by Alice measuring $A_{a|x}$ and Bob measuring $A^T_{a|x}$ allow us to estimate the lower bound for the steering robustness and to certify the steering dimensionality via Eq.~\eqref{eqn:ceiling}.
The expectation values in Eq.~\eqref{eqn:steering-ineq} are estimated via the normalised coincidence count rates for complete MUB measurements
\begin{align}\label{eqn:numer}
  \Tr\big[(A_{a|x}\otimes A^T_{b|x})\rho_{AB}\big] = \frac{N_{ab|x}}{\sum\limits_{ab}N_{ab|x}},
\end{align}
where $N_{ab|x}$ is the coincidence count rate obtained when Alice measures the projector outcome $a$ in basis $x$ and Bob the outcome $b$ in the same basis.
The number of counts $\sum_{ab}N_{ab|x}$, when averaged over all bases except the computational one, is 1596 in dimension 5 and 1876 in dimension 31.\\

\subsection{Application to other experimental platforms}
\label{app:platforms}

\begin{figure}[ht!]
  \centering
  \includegraphics[width=\columnwidth]{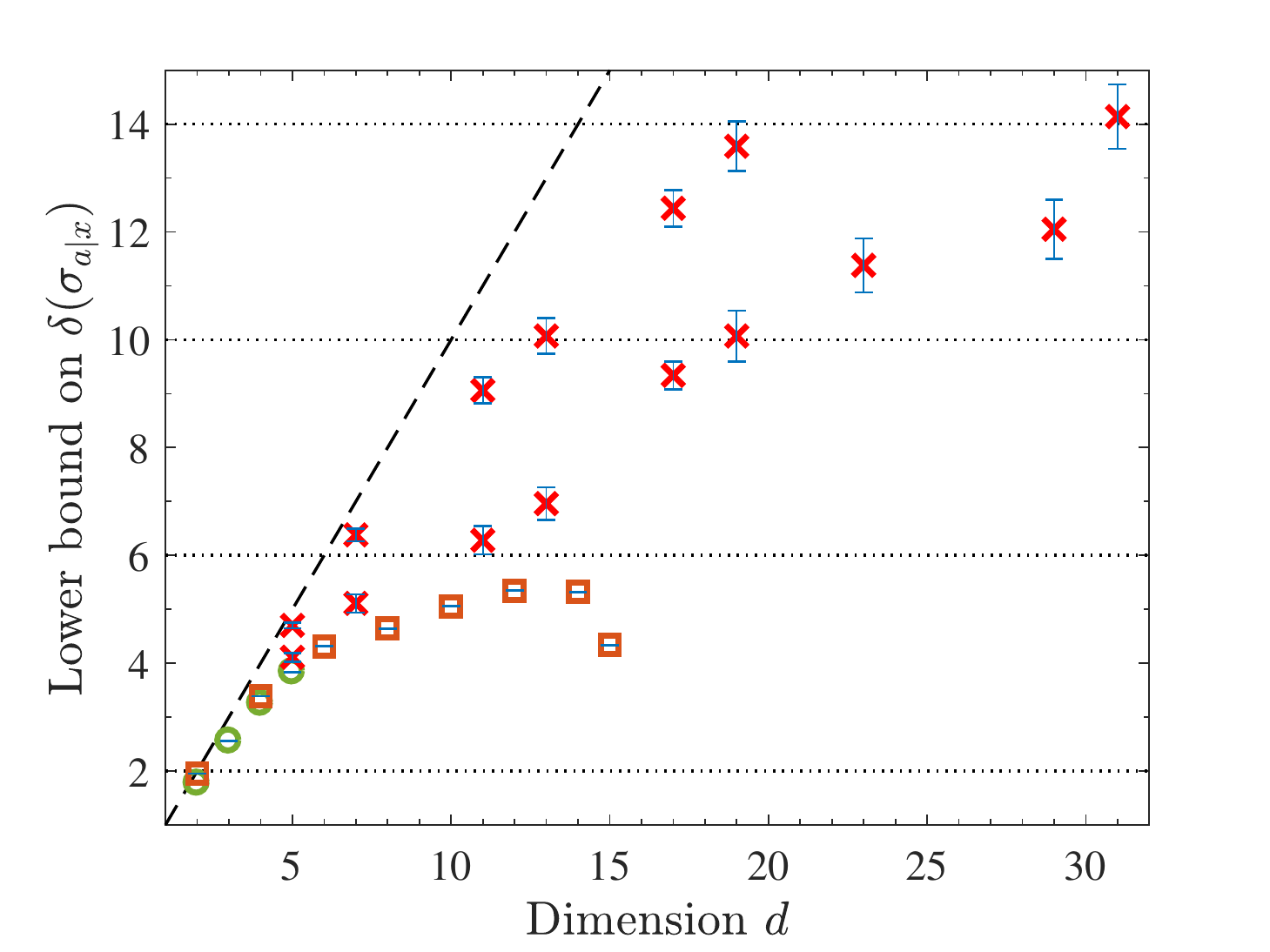}
  \caption{
    Comparison of the dimensions certified in high-dimensional two-setting quantum steering experiments.
    Green circles corresponds to the data of Ref.~\cite{ZWLZ18} based on OAM, orange squares to that of Ref.~\cite{WPD+18} with path-entanglement, and red crosses to our minimum and maximum results in prime dimensions (see Table~\ref{tab:reshigh}).
    The dashed line is $\delta(\sigma_{a|x})=d$ and errors represent one standard deviation.
  }
  \label{fig:comparison}
\end{figure}

The method presented here for testing genuine high-dimensional steering can be readily applied to other experimental platforms, such as path-entanglement in integrated chips and orbital angular momentum (OAM) entanglement.
The main ingredient needed to place a lower bound on the dimensionality of steering (and hence entanglement) is the steering robustness.
Our approach can thus be directly applied to experimental data featuring a pair of measurements on the untrusted side (Alice).
In Fig.~\ref{fig:comparison} we analyse within our framework results obtained in Refs~\cite{ZWLZ18,WPD+18} and compare them with ours.
These experiments allow one to certify genuine 6-dimensional steering, while the present results go up to 15-dimensional entanglement.

\end{document}